\documentclass[aps,twocolumn,amsmath,letterpaper]{revtex4}
\usepackage{amsmath}
\usepackage{amssymb}
\usepackage{graphicx}
\usepackage{array}
\usepackage{hhline}
\usepackage{longtable}

\begin{document}

    \title{Interface-Roughening Phase Diagram of the Three-Dimensional Ising Model for

     All Interaction Anisotropies from Hard-Spin Mean-Field Theory}

    \author{Tolga \c{C}a\u{g}lar$^{1,2}$ and A. Nihat Berker$^{1,3}$}
    \affiliation{$^1$Faculty of Engineering and Natural Sciences, Sabanc{\i} University, Orhanl{\i}, Tuzla 34956, Istanbul, Turkey,}
    \affiliation{$^2$Department of Physics, Ko\c{c} University, Sar\i yer 34450, Istanbul, Turkey,}
    \affiliation{$^3$Department of Physics, Massachusetts Institute of Technology, Cambridge, Massachusetts 02139, U.S.A.}

\begin{abstract}
The roughening phase diagram of the $d=3$ Ising model with
uniaxially anisotropic interactions is calculated for the entire
range of anisotropy, from decoupled planes to the isotropic model to
the solid-on-solid model, using hard-spin mean-field theory.  The
phase diagram contains the line of ordering phase transitions and,
at lower temperatures, the line of roughening phase transitions,
where the interface between ordered domains roughens. Upon
increasing the anisotropy, roughening transition temperatures settle
after the isotropic case, whereas the ordering transition
temperature increases to infinity. The calculation is repeated for
the $d=2$ Ising model for the full range of anisotropy, yielding no
roughening transition.

PACS numbers: 68.35.Ct, 05.50.+q, 64.60.De, 75.60.Ch

\end{abstract}

    \maketitle

\section{Introduction}

The ordering phase transition in a crystal precipitates the
formation of macroscopic domains, differently ordered with respect
to each other.  The interface between such domains incorporates
static and dynamic phenomena of fundamental and applied importance.
Of singular importance is the occurrence of yet another phase
transition, distinct from the ordering phase transition, which is
the interface roughening phase transition.\cite{ChuiWeeks, Swendsen}
The roughening phase transition is well-studied with the
three-dimensional Ising model, in the so-called solid-on-solid
limit, in which the interactions along one spatial direction ($z$)
are taken to infinite strength, while the interactions along the $x$
and $y$ spatial directions remain finite.  In this case, due to the
infinite interactions, the ordering phase transition moves to
infinite temperature and is not observed.  A study of the system
with finite interactions, where both ordering and roughening phase
transitions should distinctly be observed, had not been done.

\begin{figure}[th!]
\centering
\includegraphics*[scale=1]{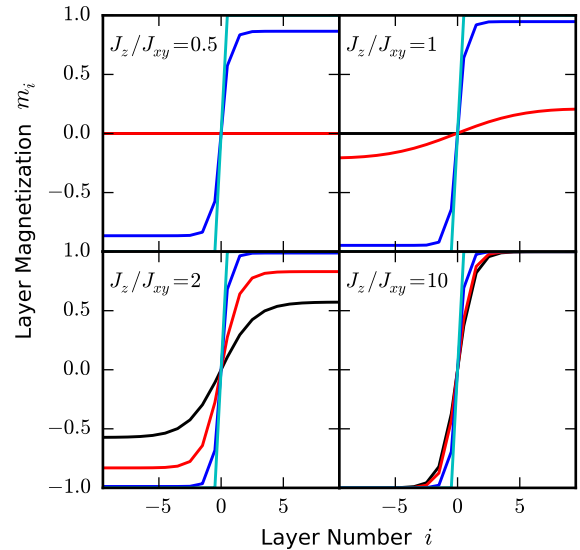}
\caption{(Color online) For the $d=3$ anisotropic Ising model,
magnetizations $m_i$ versus $xy$ layer-number $i$ curves for
different temperatures $1/J_{xy}$.  Each panel shows results for the
indicated anisotropy $J_z/J_{xy}$.  The curves in each panel, with
decreasing sharpness, are for temperatures $1/J_{xy}=1,3,5,6$.  In
the upper panels, the high-temperature curves coincide with the
horizontal line $m_i=0$.}
\end{figure}

\begin{figure}[th!]
\centering
\includegraphics*[scale=1]{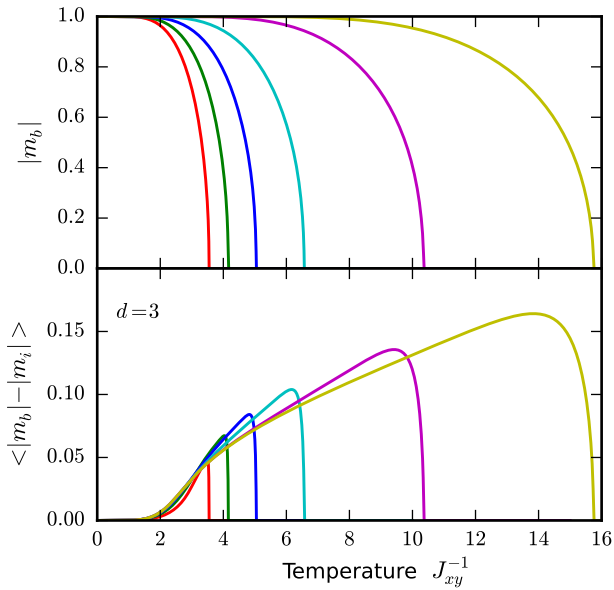}
\caption{(Color online) Local magnetization data for the $d=3$
anisotropic Ising model.  The curves, starting from the
high-temperature side, are for anisotropies
$J_z/J_{xy}=10,5,2,1,0.5,0.2\,$.  Upper panel:  Magnetization
absolute values $|m_b|$ away from the interface as a function of
temperature $1/J_{xy}$, for different values of the anisotropy
$J_z/J_{xy}$. Lower panel:  The deviation $|m_b|-|m_i|$ averaged
over the system versus temperature $1/J_{xy}$ for different
anisotropies $J_z/J_{xy}$. This averaged deviation vanishes when the
interface is smooth. Note the qualitatively different
low-temperature behavior in the $d=2$ case shown in Fig. 4.}
\end{figure}

In our current study, hard-spin mean-field theory \cite{Netz1,
Netz3}, which has been qualitatively and quantitatively successful
in frustrated and unfrustrated magnetic ordering problems
\cite{Netz1, Netz3, Banavar, Netz2, Netz4, Netz5, Berker,
Kabakcioglu2, McKayAbs, Akguc, McKay2, Monroe, Pelizzola,
Kabakcioglu, Kaya, Yucesoy}, is used to study ordering and
roughening phase transitions in the three-dimensional ($d=3$) Ising
model for the entire range of interaction anisotropies, continuously
from the solid-on-solid limit to the isotropic system to the
weakly-coupled-planes limit.  The phase diagram is obtained in the
temperature and interaction anisotropy variables, with separate
curves of ordering and roughening phase boundaries.  The method,
when applied to the anisotropic $d=2$ Ising model, yields the lack
of roughening phase transition.

\section{Hard-Spin Mean-Field Theory}

Hard-spin mean-field theory has been introduced as a self-consistent
theory that conserves the hard-spin $(|s|=1)$ condition,
indispensable to the study of frustrated systems.\cite{Netz1, Netz3}
This method is almost as simply implemented as usual mean-field
theory, but brings considerable qualitative and quantitative
improvements. Hard-spin mean-field theory has yielded, for example,
the lack of order in the undiluted zero-field triangular-lattice
antiferromagnetic Ising model and the ordering that occurs either
when a uniform magnetic field is applied to the system, giving a
quantitatively accurate phase diagram in the temperature versus
magnetic field variables \cite{Netz1, Netz3, Banavar, Netz2, Berker,
Kabakcioglu2}, or when the system is sublattice-wise quench-diluted
\cite{Kaya}. Hard-spin mean-field theory has also been successfully
applied to complicated systems that exhibit a variety of ordering
behaviors, such as three-dimensionally stacked frustrated systems
\cite{Netz1, Netz4}, higher-spin systems \cite{Netz5}, and
hysteretic $d=3$ spin-glasses \cite{Yucesoy}. Furthermore, hard-spin
mean-field theory shows qualitative and quantitative effectiveness
for unfrustrated systems as well, such as being dimensionally
discriminating by yielding the no-transition of $d=1$ and improved
transition temperatures in $d=2$ and $d=3$.\cite{Banavar, Yucesoy}

We have therefore applied hard-spin mean-field theory to the global
study of the roughening transition in the anisotropic $d=3$ Ising
model.  [We have also found that no roughening phase transition is
seen in $d=2$ (Sec.IV)].  The uniaxially anisotropic $d=3$ Ising
model is defined by the Hamiltonian
\begin{equation}
-\beta \mathcal{H} = J_{xy}\sum_{\langle ij\rangle}^{xy}s_is_j +
J_{z}\sum_{\langle ij\rangle}^zs_is_j, \label{hamilton}
\end{equation}
\noindent where, at each site $i$ of a $d=3$ cubic lattice with
periodic boundary conditions, $s_i = \pm1$. The first sum is over
nearest-neighbor pairs of sites along the $x$ and $y$ spatial
directions and the second sum is over the nearest-neighbor pairs of
sites along the $z$ spatial direction. The interactions are
ferromagnetic, $J_{xy},J_z>0$, except for the interaction between
two of the $xy$ planes, which has the same magnitude as the other
$J_z$ interactions, but is antiferromagnetic: $J_z^{A}=-J_z<0$. This
choice is made in order to induce an interface when the system is
ordered. (An alternate approach would have been to use a system
without periodic boundary conditions along the $z$ direction, but
with oppositely pinned spins at each edge.  However, this would have
introduced a surface effect at the pinned edges, modifying the
magnetization deviations which would thereby not exclusively reflect
the spreading of the interface.)

For this system, the self-consistent equation of  hard-spin
mean-field theory is
\begin{equation}
m_i =
\sum_{\{s_j\}}\left[\left(\prod_{j}P\left(m_j,s_j\right)\right)\times\tanh\left(\sum_jJ_{ij}s_j\right)\right],
\label{iter}
\end{equation}
\noindent where the last sum is over the sites $j$ that are nearest
neighbor to site $i$ and the first sum is over all states $\{s_j\}$
of the spins at these nearest-neighbor sites.  In Eq.(2),
\begin{equation}
P\left(m_j,s_j\right) = \frac{1}{2}\left(1 + m_j\*s_j\right)
\end{equation}
\noindent is, for local magnetization $m_j$ at site $j$, the
probability of having the spin value of $s_j$.  The coupled Eqs. (2)
are solved numerically for a 20x20x20 cubic system with periodic
boundary conditions, by iteration: A set of magnetizations is
substituted into the right-hand side of Eqs. (2), to obtain a new
set of magnetizations from the left-hand side.  This new set is then
substituted into the right-hand side, and this procedure is carried
out repeatedly, converging to stable values of the magnetizations
that is the solution of the equations. The resulting magnetization
values depend on the $z$ coordinate only.

\begin{figure}[th!]
\centering
\includegraphics*[scale=1]{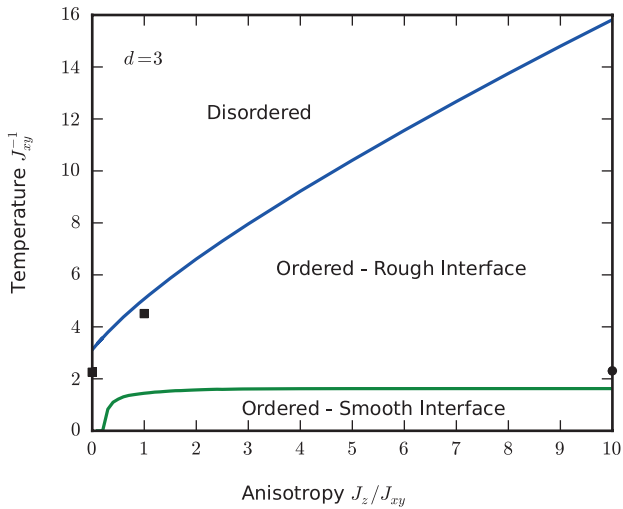}
\caption{(Color online) For the $d=3$ anisotropic Ising model, the
calculated phase diagram showing the disordered, ordered with rough
interface, and ordered with smooth interface phases.  The squares
indicate the exact ordering temperatures from duality at
$J_z/J_{xy}=0$ and from Ref. \cite{Landau} at $J_z/J_{xy}=1$.  The
circle indicates the roughening transition temperature for the
solid-on-solid limit $J_z/J_{xy}\rightarrow\infty$.\cite{Swendsen}
The roughening transition is obtained by fitting the averaged
deviation curves (lower panel of Fig. 2) within the range
$<|m_b|-|m_i|>=0.01$ to 0.04, to find the temperature at which the
averaged deviation reaches zero, meaning that the interface becomes
localized between two consecutive layers, reversing the sign of the
magnetization $m_b$ with no change in magnitude.}
\end{figure}

\section{Results: Ordering and Roughening Phase Transitions in $d=3$}

A series of curves for the magnetizations $m_i$ versus $xy$ layer
number $i$ are shown for different temperatures $1/J_{xy}$, for a
given anisotropy $J_z/J_{xy}$ in each panel of Fig. 1 .  For each
value of the anisotropy, the magnetizations $m_i$ are zero at high
temperatures and become non-zero below the ordering transition
temperature $T_C$. The ordering onset is seen in the upper panel of
Fig. 2, where the magnetization absolute values $|m_b|$ away from
the interface are plotted as a function of temperature $1/J_{xy}$,
for different values of the anisotropy $J_z/J_{xy}$.

In Fig. 1, it is also seen that, at temperatures just below $T_C$,
the interface between the $m_i\gtrless0$ domains is spread over
several layers. It is also seen that below a lower,
roughening-transition temperature $T_R$, the interface becomes
localized between two consecutive layers, reversing the sign of the
magnetization $m_i$ with no change in magnitude. This onset is best
seen in the lower panel of Fig. 2, where the deviation $|m_b|-|m_i|$
averaged over the system is plotted as a function of temperature
$1/J_{xy}$ for different anisotropies $J_z/J_{xy}$.

\begin{figure}[th!]
\centering
\includegraphics*[scale=1]{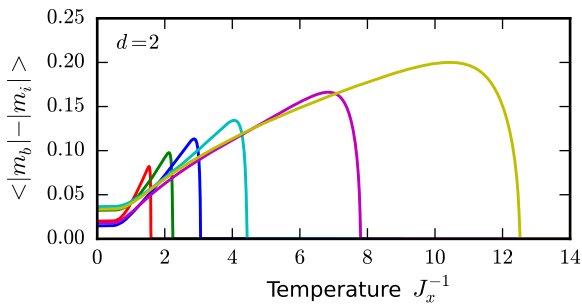}
\caption{(Color online) For the $d=2$ anisotropic Ising model, the
deviation $|m_b|-|m_i|$ averaged over the system versus temperature
$1/J_{xy}$ for different anisotropies $J_z/J_{xy}$.  The curves,
starting from the high-temperature side, are for anisotropies
$J_z/J_{xy}=10,5,2,1,0.5,0.2\,$.  It is seen that the deviation does
not vanish, i.e., the interface does not localize, down to zero
temperature.  Thus, a qualitatively different low-temperature
behavior occurs, as compared with the $d=3$ case shown in the lower
panel of Fig. 2.}
\end{figure}

Thus, we have deduced the phase diagram, for all values of the
anisotropy $J_z/J_{xy}$ and temperature $1/J_{xy}$, as shown in Fig.
3. The roughening transition is obtained by fitting the averaged
deviation curves (lower panel of Fig. 2) within the range
$<|m_b|-|m_i|>=0.01$ to 0.04, to find the temperature at which the
averaged deviation reaches zero, meaning that the interface becomes
localized between two consecutive layers, reversing the sign of the
magnetization $m_b$ with no change in magnitude. In Fig. 3, the
ordering and roughening phase transitions occur as two separate
curves, starting in the decoupled planes ($J_z/J_{xy}=0$) limit and
scanning at finite temperature the entire range of anisotropies. The
ordering transition starts, for the decoupled planes limit
$J_z/J_{xy}=0$, at $1/J_{xy}=$3.12, to be compared with the exact
result of $1/J_{xy}=2.27$.  The ordering transition continues to
$1/J_{xy}=5.06$, to be compared with the precise \cite{Landau}
result of $1/J_{xy}=4.51$, for the isotropic case $J_z/J_{xy}=1$. In
the solid-on-solid limit ($J_z/J_{xy}\rightarrow\infty$), the
ordering boundary goes to infinite temperature.  The roughening
transition starts at $1/J_{xy}=0$ for $J_z/J_{xy}$ close to zero and
settles to a finite temperature value before the isotropic case.
Thus, the roughening transition temperature $1/J_{xy}$ is 1.45 in
the isotropic case $J_z/J_{xy}=1$ and 1.62 in the solid-on-solid
limit $J_z/J_{xy}\rightarrow\infty$, the latter to be compared with
the value of $2.30\pm 0.10$ from Ref.\cite{Swendsen}.

\section{Results: Ordering Transitions but No Roughening Transitions in $d=2$}

We have also applied our method to the anisotropic d=2 Ising model,
defined by the Hamiltonian
\begin{equation}
-\beta \mathcal{H} = J_{x}\sum_{\langle ij\rangle}^{x}s_is_j +
J_{z}\sum_{\langle ij\rangle}^zs_is_j, \label{hamilton}
\end{equation}
where, on a 20x20 square lattice with periodic boundary conditions,
the first sum is over nearest-neighbor pairs of sites along the $x$
spatial direction and the second sum is over the nearest-neighbor
pairs of sites along the only other $(z)$ spatial direction.

The ordering phase transition is observed in $d=2$ similarly to the
$d=3$ case. However, the rough interface phase continues to zero
temperature, as seen in the $|m_b| -|m_i|$ curves in Fig. 4. Thus,
no roughening phase transition occurs in $d=2$.  The corresponding
phase diagram is given in Fig. 5.  The ordering transition starts,
for the decoupled lines limit $J_z/J_{x}=0$, at $1/J_{x}=0$, as
expected for decoupled $d=1$ systems.  The ordering transition
continues to $1/J_{x}=$3.09, to be compared with the exact result of
$1/J_{x}=2.27$, for the isotropic case $J_z/J_{x}=1$. In the
$J_z/J_{x}\rightarrow\infty$ limit, the ordering boundary goes to
infinite temperature.

\begin{figure}[th!]
\centering
\includegraphics*[scale=1]{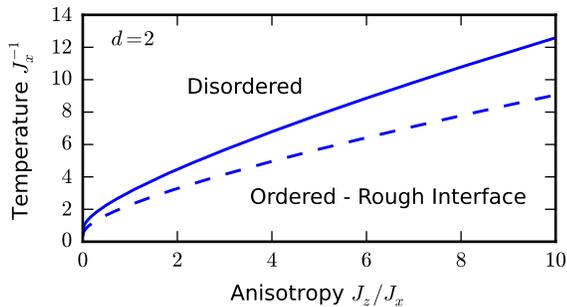}
\caption{(Color online) For the d=2 anisotropic Ising model, the
phase diagram showing the disordered phase and the ordered phase
with rough interface.  The dashed curve is the exact ordering
boundary $sinh(2J_x)sinh(2J_z)=1$ obtained from duality.  No ordered
phase with smooth interface is found.}
\end{figure}

\section{Conclusion}

It seen that hard-spin mean-field theory yields a complete picture
of the ordering and roughening phase transitions for the isotropic
and anisotropic Ising models, in spatial dimensions $d=3$ and 2.
This result attests to the microscopic efficacy of the model. Future
works, such as the effects of uncorrelated and correlated (aerogel
\cite{Chan, Falicov}) frozen impurities on the roughening
transitions, are planned.

\begin{acknowledgments}
Support by the Alexander von Humboldt Foundation, the Scientific and
Technological Research Council of Turkey (T\"UB\.ITAK), and the
Academy of Sciences of Turkey is gratefully acknowledged.
\end{acknowledgments}


\begin{thebibliography}{}

\bibitem{ChuiWeeks} S.T. Chui and J.D. Weeks, Phys. Rev. {\bf 14}, 4978 (1976).
\bibitem{Swendsen} R.H. Swendsen, Phys.Rev. B {\bf 15}, 5421 (1977).
\bibitem{Netz1} R.R. Netz and A.N. Berker, Phys. Rev. Lett. {\bf 66}, 377 (1991).
\bibitem{Netz3} R.R. Netz and A.N. Berker, J. Appl. Phys. {\bf 70}, 6074 (1991).
\bibitem{Banavar} J.R. Banavar, M. Cieplak, and A. Maritan, Phys. Rev. Lett. {\bf 67}, 1807 (1991).
\bibitem{Netz2} R.R. Netz and A.N. Berker, Phys. Rev. Lett. {\bf 67}, 1808 (1991).
\bibitem{Netz4} R.R. Netz, Phys. Rev. B {\bf 46}, 1209 (1992).
\bibitem{Netz5} R.R. Netz, Phys. Rev. B {\bf 48}, 16113 (1993).
\bibitem{Berker} A.N. Berker, A. Kabak\c{c}\i o\u{g}lu, R.R. Netz, and M.C. Yalab\i k, Turk.
J. Phys. {\bf 18}, 354 (1994).
\bibitem{Kabakcioglu2} A. Kabak\c{c}\i o\u{g}lu, A.N. Berker, and M.C. Yalab\i k, Phys. Rev. E {\bf 49}, 2680
(1994).
\bibitem{McKayAbs} E.A. Ames and S.R. McKay, J. Appl. Phys. {\bf 76}, 6197 (1994).
\bibitem{Akguc} G.B. Akg\"u\c{c} and M.C. Yalab\i k, Phys. Rev. E {\bf 51}, 2636 (1995).
\bibitem{McKay2} J.E. Tesiero and S.R. McKay, J. Appl. Phys. {\bf 79}, 6146,
(1996).
\bibitem{Monroe} J.L. Monroe, Phys. Lett. A {\bf 230}, 111 (1997).
\bibitem{Pelizzola} A. Pelizzola and M. Pretti, Phys. Rev. B {\bf 60}, 10134 (1999).
\bibitem{Kabakcioglu} A. Kabak\c{c}\i o\u{g}lu, Phys. Rev. E {\bf 61}, 3366 (2000).
\bibitem{Kaya} H. Kaya and A.N. Berker, Phys. Rev. E {\bf 62}, R1469 (2000); also see M. D. Robinson, D. P. Feldman, and S. R. McKay,
Chaos {\bf 21}, 037114 (2011).
\bibitem{Yucesoy} B. Y\"{u}cesoy and A. N. Berker, Phys. Rev. B {\bf 76}, 014417
(2007).
\bibitem{Landau} A.M. Ferrenberg and D.P. Landau, Phys. Rev. B {\bf 44}, 5081 (1991).
\bibitem{Chan} S.B. Kim, J. Ma, and M.H.W. Chan, Phys. Rev. Lett. {\bf 71}, 2268 (1993).
\bibitem{Falicov} A. Falicov and A.N. Berker, Phys. Rev. Lett. {\bf 74}, 426 (1995).
\end{thebibliography}
\end{document}